\documentclass[prl,twocolumn,superscriptaddress,10pt]{revtex4-1}
\usepackage{graphicx}
\usepackage{dcolumn}
\usepackage{color}
\usepackage{times}
\usepackage{bm}
\usepackage{amssymb}
\usepackage{amsmath}
\usepackage{epsfig}
\usepackage{epstopdf}
\usepackage{dsfont}
\usepackage{subfigure}
\usepackage[colorlinks=true,citecolor=blue, linkcolor=blue]{hyperref}

\begin{document}
\renewcommand{\thefootnote}{\fnsymbol {footnote}}

\title{Towards Heisenberg limit without critical slowing down via quantum reinforcement learning}

\author{Hang Xu}
\affiliation{State Key Laboratory of Advanced Optical Communication Systems and Networks, Institute for Quantum Sensing and Information Processing, Shanghai Jiao Tong University, Shanghai 200240, P.R.  China}%

 \author{Tailong Xiao}%
\email{tailong\_shaw@sjtu.edu.cn}
\affiliation{State Key Laboratory of Advanced Optical Communication Systems and Networks, Institute for Quantum Sensing and Information Processing, Shanghai Jiao Tong University, Shanghai 200240, P.R.  China}%
 \affiliation{Hefei National Laboratory, Hefei, 230088, P.R. China}
\affiliation{Shanghai Research Center for Quantum Sciences, Shanghai, 201315, P.R. China}

 \author{Jingzheng Huang}%
\affiliation{State Key Laboratory of Advanced Optical Communication Systems and Networks, Institute for Quantum Sensing and Information Processing, Shanghai Jiao Tong University, Shanghai 200240, P.R.  China}%
 \affiliation{Hefei National Laboratory, Hefei, 230088, P.R. China}
\affiliation{Shanghai Research Center for Quantum Sciences, Shanghai, 201315, P.R. China}

\author{Ming He}
\affiliation{AI Lab, Lenovo Research, Beijing 100094, P.R.  China}%

\author{Jianping Fan}
\affiliation{AI Lab, Lenovo Research, Beijing 100094, P.R.  China}%

 \author{Guihua Zeng}%
 \email{ghzeng@sjtu.edu.cn}
\affiliation{State Key Laboratory of Advanced Optical Communication Systems and Networks, Institute for Quantum Sensing and Information Processing, Shanghai Jiao Tong University, Shanghai 200240, P.R.  China}%
 \affiliation{Hefei National Laboratory, Hefei, 230088, P.R. China}
\affiliation{Shanghai Research Center for Quantum Sciences, Shanghai, 201315, P.R. China}

\date{\today}

\begin{abstract}{
Critical ground states of quantum many-body systems have emerged as vital resources for quantum-enhanced sensing. Traditional methods to prepare these states often rely on adiabatic evolution, which may diminish the quantum sensing advantage. In this work, we propose a quantum reinforcement learning (QRL)-enhanced critical sensing protocol for quantum many-body systems with exotic phase diagrams. Starting from product states and utilizing QRL-discovered gate sequences, we explore sensing accuracy in the presence of unknown external magnetic fields, covering both local and global regimes. Our results demonstrate that QRL-learned sequences reach the finite quantum speed limit and generalize effectively across systems of arbitrary size, ensuring accuracy regardless of preparation time. This method can robustly achieve Heisenberg and super-Heisenberg limits, even in noisy environments with practical Pauli measurements. Our study highlights the efficacy of QRL in enabling precise quantum state preparation, thereby advancing scalable, high-accuracy quantum critical sensing.
}

\end{abstract}

\maketitle

\textit{Introduction.}---The accuracy of classical sensors increases linearly with the amount of resources used,
$N$, following the standard quantum limit.  Quantum systems are more sensitive to their environment due to coherence and entanglement.
Quantum criticality \cite{critcal1,critcal2} can also serve as an indicator to improve sensing accuracy \cite{critcalsensing1,xiao2022parameter,sensitivity1,sensitivity2,sensitivity3,weak1,ptUhlmann,multiUhlmann,multiesti}. These probes exploit various critical phenomena, including second-order \cite{secondorder3,secondorder4}, topological \cite{topo1,topo2,topo3}, dissipation \cite{dissipation1,dissipation2,dissipation3,dissipation4,crystals1,dpUhlmann}, and measurement-induced phase transitions \cite{measpahse}.
Generally, many-body critical sensing can be categorized into based on ground state phase transitions in equilibrium conditions and dynamic phase transitions via time evolution.
In the former case, near the phase transition point, ground states transition rapidly from one phase to another, resulting in a much faster rate of property change than within a single quantum phase region.
The rapid change of quantum state property facilitates surpassing the standard limits of classical sensing systems, enabling the achievement of the Heisenberg limit or even beyond \cite{superHeisenberg2,localization1}.

The complexity of quantum dynamics grows exponentially with system sizes, making the time resources needed to prepare the critical ground state by adiabatic evolution \cite{adiabatic1,counterdiabatic1} grow with the system sizes, which is referred to as critical slowing down \cite{stronglimit,sachdev1999quantum}.
If the total time of the sensing scheme is fixed fairly, the increase in preparation time with the system size leads to a decrease in the number of performable measurements with the size, which ultimately leads to a degradation of the accuracy limit in the Cram\'{e}r-Rao boundary.
It has been proved that using adiabatic evolution, quantum critical sensing cannot reach Heisenberg limit \cite{gietka2021adiabatic}.
Although numerous machine learning methods have been proposed to address this exponential growth challenge \cite{mlsensing1,mlsensing2,mlsensing3,mlsensing4,mlsensing5,mlsensing6, qdqn, xiao2023quantum,rl_cd}, they still necessitate a large number of quantum measurements or rely on unrealistic assumptions \cite{zhu2024controlling}.

In this work, we propose a quantum reinforcement learning-enhanced critical sensing (QRLCS) protocol, which is based on the criticality of many-body systems under equilibrium conditions, and achieves sensing accuracy at the Heisenberg or super-Heisenberg limit without encountering critical slowing down.
We examine the quantum Ising and XY spin chains as models for probing magnetic fields. By employing QRL with terms occurring in the adiabatic gauge potential as generators of gate controls, we find that the discovered gate control sequences can fast prepare the critical ground states away from the adiabatic regime and reaches the finite quantum speed limit. These ground states are then utilized as probes to attain the desired accuracy, exhibiting robustness to noise under practical single-site Pauli measurements.
Unlike conventional many-body sensing schemes, the preparation time required for our protocol is fixed and the learned gate sequences can generalize to arbitrary system sizes, thus realizing a true Heisenberg or super-Heisenberg limit.

\textit{Critical sensing.}---A complete quantum parameter estimation process consists of: (1) preparing an optimal probe $\rho_0$, (2) allowing the probe to interact with the environment to encode unknown parameters $\bm{h}$, yielding $\rho_{\bm{h}}$, (3) performing repeated measurements on the encoded probe to collect statistics $\langle \mathcal{O}_{\bm{h}}\rangle = \text{Tr} [\rho_{\bm{h}} \mathcal{O}]$, and (4) using these statistics to estimate $\bm{h}$. The variance of the estimator is subject to the Cram\'{e}r-Rao inequality \cite{CramerRao1}:
\begin{equation}
{\mathop{\rm cov}} (\bm{h}) \ge {M^{ - 1}}{F_C}{(\bm{h})^{ - 1}} \ge {M^{ - 1}}{F_Q}{(\bm{h})^{ - 1}},\label{two-inequality}
\end{equation}
where ${\mathop{\rm cov}} (\bm{h})$ is the covariance matrix, ${F_C}$ and ${F_Q}$ represent the classical Fisher information (CFI) and quantum Fisher information (QFI) matrices, respectively.
$M = \mathcal{T}{\rm{/(}}{{t}_p}{\rm{ + }}\tau_m {\rm{)}}$ is the number of measurements, where $\mathcal{T}$ is the total time required for the entire sensing protocol, $t_p$ is the probe preparation time, and $\tau_m$ is the interrogation time of single shot.

Generally, the precision limit of critical sensing, without considering the preparation time of the critical state, is given by $F_Q \propto N^{2/d\nu}$ \cite{rams2018limits}, where $\nu$ is the correlation length critical exponent, $d$ is the spatial dimension and $N=L^d$ denotes the number of spins.
Under adiabatic limit \cite{knysh2016zero}, the time to prepare the instantaneous ground state is at least  scaled as $t_p \propto N^{\mu/d}$ \cite{schwandt2009quantum, rams2018limits}, where $\mu$ is the dynamical critical exponent.
This means that the number of measurements of the critical probe is dependent on the system size.
Considering the long preparation time of the many-body ground state, a natural assumption is ${t_p} \gg \tau_m $. Then we have $M \approx \mathcal{T}/{t_p}$.
In order to fairly compare the accuracy of probes of different sizes, Eq.~(\ref{two-inequality}) is rewritten as
\begin{equation}
{\mathop{\rm cov}} (\bm{h}) \ge {\mathcal{T}^{ - 1}}{\mathcal{F_{C}}^{ - 1}} \ge {\mathcal{T}^{ - 1}}{\mathcal{F_{Q}}^{ - 1}},
\end{equation}
where $\mathcal{F_{Q,C}} = {F_{Q,C}}/{t_p}$ is defined as the time-factorized QFI (CFI).
As for 1-dimensional Ising model, we have $\mu=1, \nu=1$ \cite{rams2018limits,stronglimit}. Thus, the time-factorized QFI becomes $\mathcal{F_Q} \propto N$, degrading from Heisenberg limit (${F_Q} \propto {N^2}$) to the standard quantum limit. To overcome this issue, it is crucial to develop schemes for preparing critical ground states where the required time $t_p$ is independent of $N$, allowing the recovery of the Heisenberg limit for the critical sensing model.

\begin{figure}[htbp]
  \centering
\includegraphics[width=0.9\linewidth]{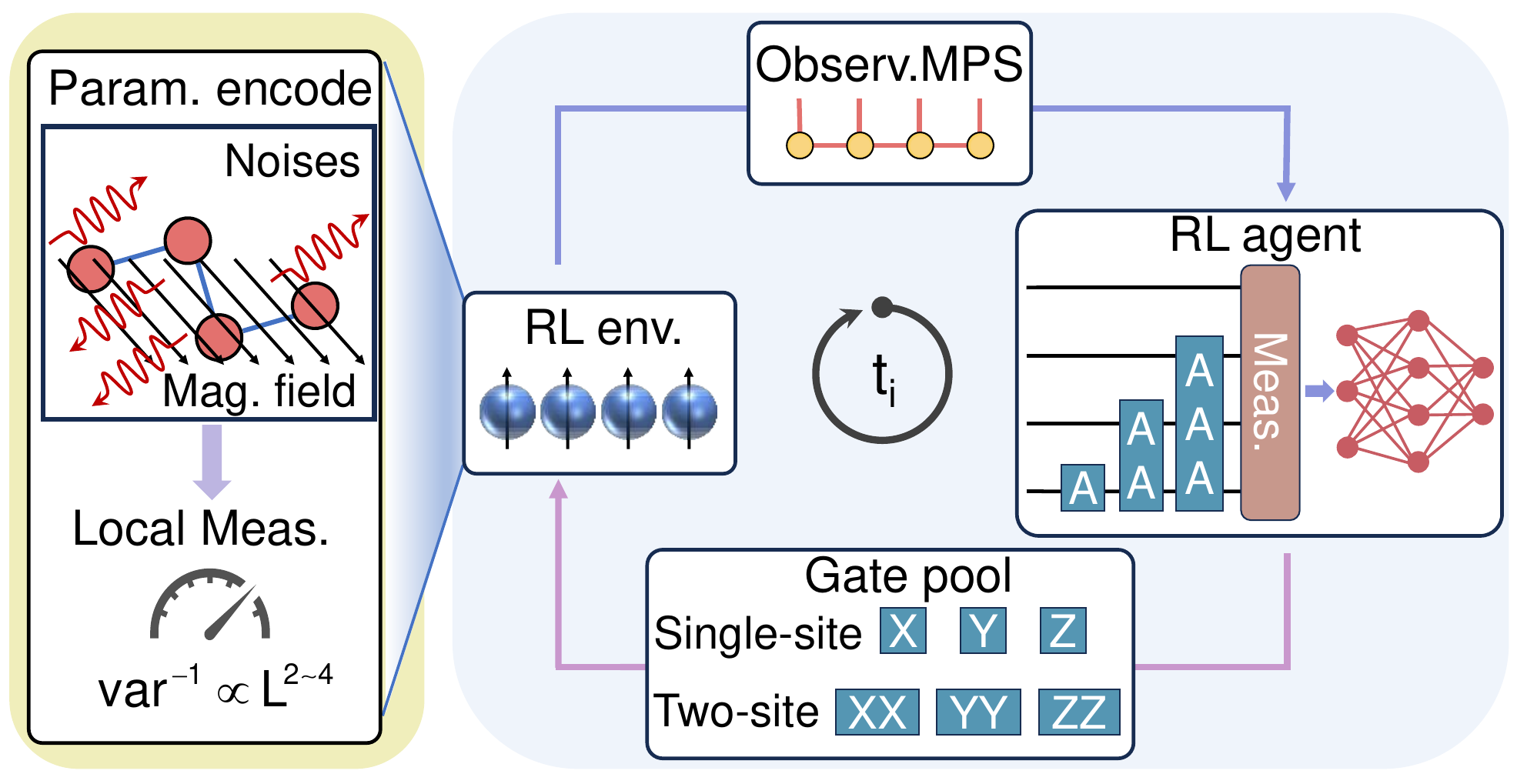}
  \caption{Schematic of the quantum reinforcement learning critical sensing (QRLCS) protocol.}
  \label{frame}
\end{figure}

\textit{The QRLCS protocol.}--- The QRLCS protocol comprises two stages, illustrated in Fig.~\ref{frame}. Initially, a QRL agent learns to prepare the critical ground state of a many-body system, which serves as the sensing probe. Subsequently, the probe is immersed in an environment characterized by unknown parameters, and the controllable field is adjusted to optimize the probe's sensitivity to these parameters, facilitating high-precision measurements.

As shown in Fig.~\ref{frame}, QRL preparation of a quantum state can be modeled as a Markov decision process. The quantum agent consisting of a quantum circuit and a neural network takes as input of the state of the many-body system as an observation $s\in \mathcal{S}$ from the environment, where the environment is the many-body system, the observation is the quantum state of many-body system, and $\mathcal{S}$ is the set of all possible observations \cite{suppl1}.
Note that when quantum memory is available, one can directly load the quantum state as the input of the agent. Otherwise, using matrix product state \cite{mps0,mps1,mps2} is also efficient to characterize the low-entangled many-body states. Then, the quantum agent selects an action $a\in \mathcal{A}$ from the action space $\mathcal{A}$ to the environment based on current observation, and so on until the many-body system in the environment evolves to the target critical ground state. The ultimate goal of QRL is to maximize the accumulated reward $\mathcal{R} = \sum_{i=1}^T \eta^i r_i(s,a) $ when the quantum agent interacts with the environment with a fixed time steps $T$, where $\eta$ is the discount rate, $r_i\propto{|\langle \psi_i|\psi_{crt}\rangle |^2}$ denotes the reward function. The quantum policy network $\pi_\theta(a|s)$ parameterized by $\theta$ can be optimized with gradient ascent method such as Adam \cite{kingma2014adam}, i.e. $\Delta \theta \sim \nabla_\theta \bar{\mathcal{R}}$ where $\bar{\mathcal{R}}$ denotes the average return over numerous episodes.

Considering the ease of implementation in current quantum devices such as  superconducting qubits, neutral atoms, and trapped ions \cite{suppl2},
the control actions involve only single and two-site Pauli generators from the adiabatic gauge potential \cite{rl_cd} denoted as
\begin{equation}
    \begin{split}
           a\in  \mathcal{A} =& \biggl\{  \sum_i \sigma^x_i, \sum_i \sigma^y_i, \sum_i \sigma^z_i,  \\
     & \sum_i \sigma^x_i\sigma^x_{i+1}, \sum_i \sigma^y_i\sigma^y_{i+1},\sum_i \sigma^z_i\sigma^z_{i+1} \biggl \},
    \end{split}
\end{equation}
where $\sigma^x,\sigma^y,\sigma^z$ are the Pauli $x,y,z$ operators, respectively.
The gauge potential is derived from the adiabatic shortcut technique, which is used to realize non-adiabatic transfers from the ground state to the target ground state. Typically the gauge potentials for spin chain models are linear combinations of single and two-site Pauli operators \cite{gauge1}.
The quantum agent is trained to prepare the target ground state $|\psi_{crt}\rangle$ from a product state $|\psi_0\rangle $ by using $|\psi_{crt}\rangle = \prod_{k=1}^D\exp(-i \tau a_k) |\psi_0\rangle$, where $D$ denotes the depth of the quantum circuit, $a_k$ denotes the $k$th action layer and $\tau_g$ is the fixed gate duration. The preparation time is thus $t_p=D\tau_g$. One can also optimize the gate duration to further reduce the circuit depth. The cardinality of the search space (the set of all possible combinations of gates in action space $\mathcal{A}$) to prepare the critical target state is $|\mathcal{A}|^{D}$, exponentially increased with the depth and presenting a challenging discrete combinatorial optimization problem; therefore, a brute-force search algorithm becomes
prohibitively expensive. QRL can increase the number of independent discrete control degrees of freedom to $|\mathcal{A}|$, which enables us to reach larger parts of the Hilbert space with smaller circuit depth.
The simple action space $\mathcal{A}$ ensures the experimental feasibility of QRL.

After preparing the critical probes, we adjust the control field to maximize the sensitivity of the probes to the unknown parameters.
In general, the unknown parameter is not at a critical point, indicating that the probe is not sensitive to it, but the parameter can be moved to the critical point by adding an additional control field ${B}$, i.e., ${B} + {h} = {B_{crit}}$ to maximize the QFI.
We usually assume that \textit{a priori} information about $h$ is provided, otherwise, adaptive feedback methods are required to achieve the adjustment \cite{salvia2023critical}.
Here, suppose the unknown parameter is distributed within \textit{a priori} interval (e.g., in global sensing scenarios), the system's sensing accuracy can be quantified by the average QFI-based uncertainty, i.e., $\mathcal{K_Q} = \int_{\Delta {h}} {f({h}){\mathcal{F_Q}}{{({h})}^{ - 1}}} \text{d}{h}$,
where $\Delta {h} \in [h^{\min },h^{\max }]$ is the prior distribution interval and $f({h})$ is the prior distribution function. When the center of the sensing parameter distribution interval is at the critical point, i.e., ${B} + (h^{\min } + h^{\max })/2 = {B_{crit}}$, the QFI-based average uncertainty reaches the minimum.

The quantum gates from the gauge potential provide additional control directions in Hilbert space, which leads to the many-body system evolving along a nonadiabatical path to to critical ground state in a short time.
The translation invariance of the controls and ground states ensures that the gate sequences learned by QRL are equally valid for systems of different sizes.
This generalization guaranteed by physical principles ensures that the preparation time of critical probe remains consistent across different sized many-body systems (see SM).
The time-factorized QFI of QRLCS protocol can be written as ${\mathcal{F_Q}} \propto N^{2/d\nu}/{D\tau_g} $, thus avoiding critical slowing down.
The state preparation can be realized by a quantum computer, considering that the scheme learned by QRL is optimal and independent of the system size, and thus the time cost reaches a finite quantum speed limit \cite{QSL1,gauge1}.

To saturate the QFI, we usually requires geometrically non-local measurements based on symmetric logarithmic derivatives \cite{SLD,GaussianSLD}. However, considering practical constraints, we resort to single-site or two-site Pauli measurement. While the obtained CFI may not fully saturate the maximum QFI, achieving the Heisenberg or even higher scaling can also be realized.

\textit{Ising model.}---We consider the Ising spin chain system interacting with the control magnetic field $B_x$ and the measured magnetic field $h_x$ with the Hamiltonian of
\begin{equation}
H_{\text{Ising}} = \sum\limits_{i = 1}^{L - 1} {\sigma _i^z\sigma _{i + 1}^z}  + {g_x}\sum\limits_{i = 1}^L {\sigma _i^x},
 \label{Htwo}
\end{equation}
where $g_x = B_x + h_x$.
Here, we use the critical ground state (${g_x} = 1$) \cite{xyphase} as the sensing probe.
\begin{figure}[h]
  \centering
  \includegraphics[width=0.9\linewidth]{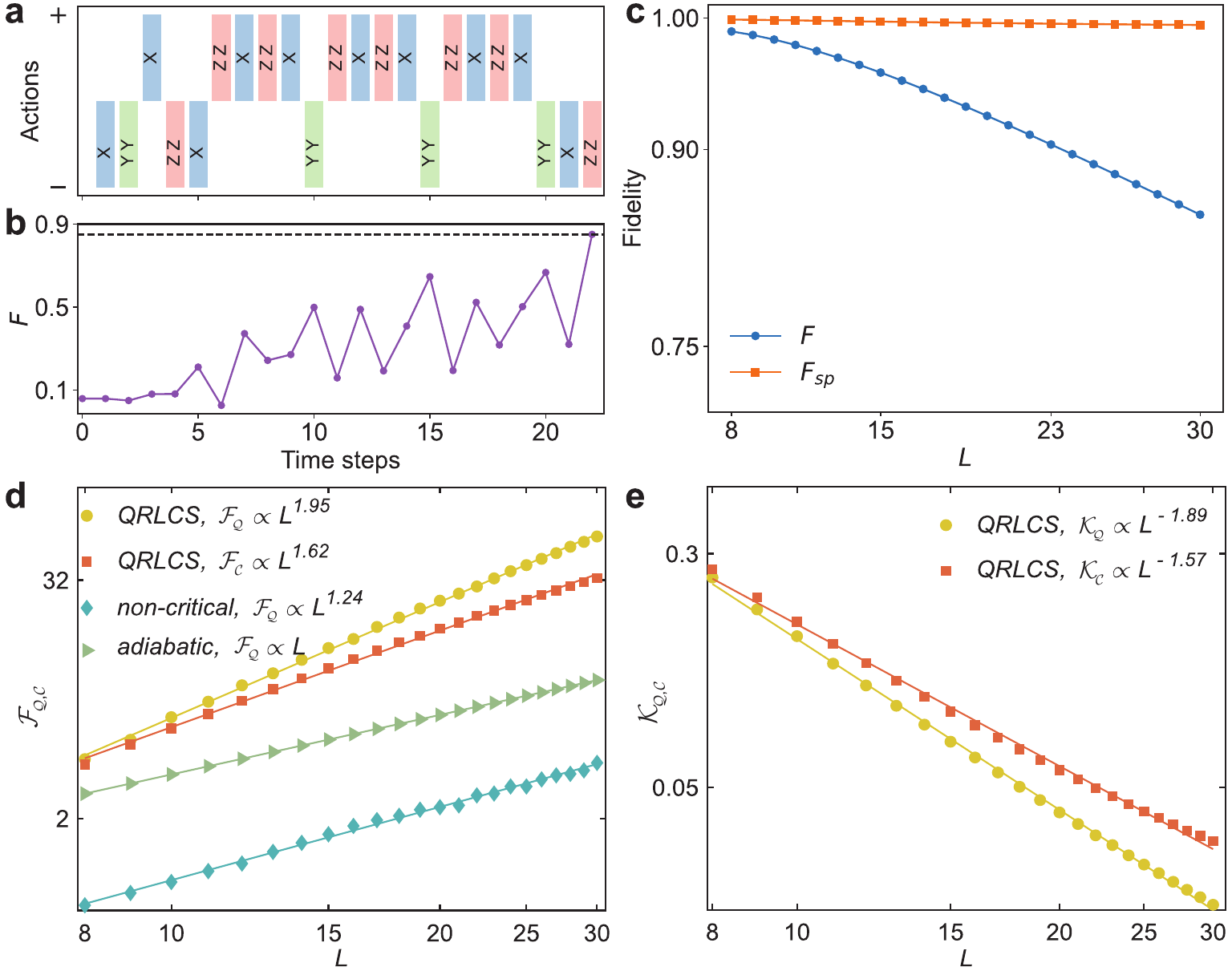}
  \caption{
  (a) Gate sequence learning by QRL, where the positive (negative) sign on the y-axis indicates the duration of the corresponding actions as positive (negative) time.
  (b) Total fidelity $F$ of the target state as a function of the number of steps (number of gates), and the total fidelity threshold $F^* = 0.85$ is marked by the black dashed line.
  (c) Fidelity of critical ground states of arbitrary spin length $L$ prepared from the action sequence in (a).
  Ising model's sensing performance of different protocols for the local (d) and global (e) case, where the probes of QRLCS are prepared from the action sequence in (a)}.
  \label{isingcontrol}
\end{figure}

During training, to ensure that the QRL model converges to learn the preparation scheme, we set the fidelity or reward function threshold to $F^* = 0.85$.
In Fig.~\ref{isingcontrol}, we depict the gate sequences  corresponding to the converged QRL scheme and the relationship between fidelity and the number of gates during the preparation of the target ground state of the Ising spin chain with $N=L=30$.
The fidelity of the target state gradually approaches to the desired value $F^*$ as the number of specific gate operations increases, where local fluctuations reveal that this is a global optimization process.
This is due to the fact that the current gate may not directly enhance the fidelity between the current state and target state, but can enhance the fidelity between the final state and target state.
Moreover, the spin ground state preparation scheme for $L<30$ can be directly generalized from the trained model ($L=30$) without retraining (see Fig.~\ref{isingcontrol}(c)).
Our QRL scheme achieves the final fidelity of 0.85, which is highly valuable for systems with $L=30$.
Of course, we can train directly on $L > 30$ and obtain gate sequences with better generalization (see SM).

Based on the critical probes already prepared by QRL, we evaluate their sensing performance.
In order to estimate an arbitrary $h_x$ with maximum accuracy, we adjust ${B_x}$ such that ${g_x}=1$ can maximize the QFI of ${h_x}$.
For a ground state $\left| {\varphi ({g_x})} \right\rangle$, its QFI is characterized by its sensitivity to the sensing parameter, i.e., ${F_{Q}} = 8(1 - f)/\delta h_x^2$,
where $f = \left| {\left\langle {{\varphi ({g_x})}}
 \mathrel{\left | {\vphantom {{\varphi ({g_x})} {\varphi ({g_x} + \delta {h_x})}}}
 \right. \kern-\nulldelimiterspace}
 {{\varphi ({g_x} + \delta {h_x})}} \right\rangle } \right|$.
Note that the probe prepared by our protocol is not a perfect critical ground state, while its effect is only a linear decay of the sensing accuracy (see SM).
Thus the true QFI of QRLCS is defined as ${F_{Q}} = 8F(1 - f)/\delta h_x^2$, where $F$ is the fidelity of the critical probe.

In Fig.~\ref{isingcontrol}(d), we present the $\mathcal{F_{Q}}$ as a function of spin length $L$
for different protocols. It is evident that the local sensing accuracy of QRLCS  almost achieves the Heisenberg limit $\mathcal{F_{Q}} \propto {L^{1.95}}$ despite using an imperfect critical ground state as its probe, underscoring the distinct advantage of our protocol. In contrast, the accuracy of the adiabatic protocol reaches, at best, the standard quantum limit $\mathcal{F_{Q}} \propto L$.
This limitation arises from the one-dimensional nearest-neighbor interaction characteristic of the spin chain \cite{stronglimit}. Moreover, the accuracy is only marginally quantum-enhanced ($\mathcal{F_{Q}} \propto {L^{1.24}}$) for probes with non-critical points.
For global sensing, we assume that ${h_x}$ is uniformly distributed over the dynamic interval.
Here, we set $\Delta {h_x}=0.1$.
In Fig.~\ref{isingcontrol}(e), we plot the function relationship between $\mathcal{K}_{Q}$ and $L$.
It can be seen that the global accuracy of our protocol reaches $\mathcal{K_{Q}} \propto {L^{-1.89}}$.
Although the dynamic range of the magnetic field from 0 to 0.1, the sensing accuracy limit of the QRLCS protocol hardly recedes.

We also analyze the CFI of the critical probe under single-site and two-site measurements (see SM).
The optimal measurement operator is $\sum\nolimits_{i = 0}^L {\sigma _i^x}$,
which makes our protocols more convenient and efficient.
In Fig.~\ref{isingcontrol}, we plot the time-factorized CFI and CFI-based average uncertainty of QRLCS varied with the spin length $L$ under measurement observable $\sum\nolimits_{i = 0}^L {\sigma _i^x}$.
The results show that the local and global sensing accuracy of our protocol can reach $\mathcal{F_{C}} \propto {L^{1.62}}$ and $\mathcal{K_{C}} \propto {L^{-1.57}}$.
Although they do not saturate the QFI, there is still a large quantum enhancement compared to adiabatic scheme.
This emphasizes the effectiveness of our protocol in the Ising model.

\textit{XY model.}---The Hamiltonian of the XY model is defined as
\begin{equation}
H_{\text{XY}} = \sum\limits_{i = 1}^{L - 1} {(\frac{{1 + \gamma }}{2}\sigma _i^x\sigma _{i + 1}^x + \frac{{1 - \gamma }}{2}\sigma _i^y\sigma _{i + 1}^y)}  + {g_z}\sum\limits_{i = 1}^L {\sigma _i^z},
\label{XYH}
\end{equation}
where $\gamma$ is the anisotropic parameter and $g_z = B_z + h_z$.
For this model, the boundary between its ferromagnetic and paramagnetic phases is the straight line $g_z = 1$ \cite{xyphase}.
Here we take the ground state in this line as the sensing probes.

\begin{figure}[h]
  \centering
  \includegraphics[width=0.9\linewidth]{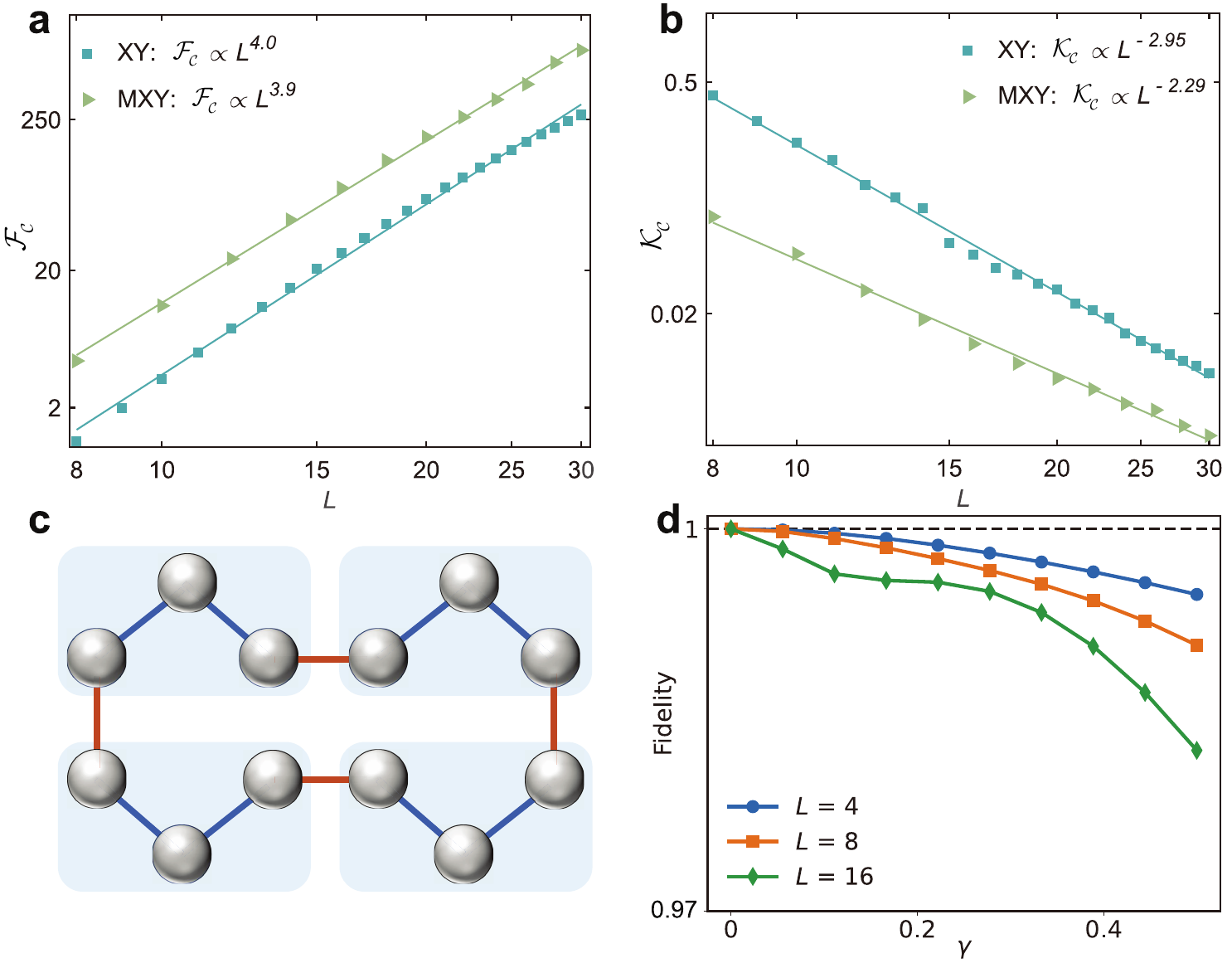}
  \caption{(M)XY model's sensing performance under measurement operator $\sum\nolimits_{i = 1}^L {{\bm{\sigma} _z}}$ of QRLCS for the local (a) and global (b) case.
  (c) Schematic of a simple MXY chain.
  (d) The fidelity between the critical ground state of the MXY model and that of the XY model.
}
  \label{XYSUM}
\end{figure}
We analyze the sensing accuracy of our protocol based on the measurement data.
The optimal measurement operator is $\sum\nolimits_{i = 0}^L {\sigma _i^z}$ (see SM).
We present in Fig.~\ref{XYSUM} the curves of the local precision characterized by the CFI and the global precision characterized by the CFI-based average uncertainty versus the spin length $L$ with optimal measurement operator, respectively.
The results show that the actual measurement accuracy of our protocol reaches the super-Heisenberg $\mathcal{F_{C}} \propto {L^{4}}$ and $\mathcal{K_{C}} \propto {L^{-2.95}}$ for the local and global cases, respectively.

Complex forms of coupling between spins increase the space for parameter encoding.
We take the modular XY (MXY) model as an example, as shown in Fig.~\ref{XYSUM}(c), where the spin chain consists of several identical modules, each consisting of a number of spins, and the coupling coefficients within the modules are different from those between the modules.
According to Ref.~\cite{MXY1}, the sensitivity interval of the ground state to the unknown parameter is larger for the model with a module length of $2$.
Therefore, we consider the case with a module length of $2$.
For this case, the critical line for the MXY model is ${g_z} \approx 0.7$.
Interestingly, the critical ground states of the MXY model are identical to those of the XY model, so there is no need to learn its preparation scheme. We illustrate this in Fig.~\ref{XYSUM}(d).

Again, we evaluate the actual accuracy of our protocol in the MXY model by using measurement statistics under the chosen measurement operator.
As shown in Fig.~\ref{XYSUM}, and we display the curves of the local and the global sensing accuracy varied with the spin length $L$ under measurement operator $\sum\nolimits_{i = 0}^L {\sigma _i^z}$, respectively.
It can be found that the actual accuracy of MXY in the local and global cases reaches $\mathcal{F_{C}} \propto {L^{3.9}}$ and $\mathcal{K_{C}} \propto {L^{-2.29}}$, respectively. This is similar to the results of the XY model, but provides higher absolute accuracy.

\textit{Noise robustness.}---We consider two common types of noise: random duration noise and cross-talk noise.
Both noise types affect the fidelity of the gates required for probe state preparation.
Random duration noise is modeled as the additional Gaussian white noise $\varepsilon$ generated over the duration of the quantum gate, obeying the distribution $\sim \mathcal{N}(0, {10}^{-2})$.
Cross-talk noise is modelled as the presence of unwanted $ZZ$ couplings in the quantum circuits performing the gate operation. In current Noisy Intermediate-Scale Quantum device that qubit error rates of the order of ${\rm{1}}{{\rm{0}}^{{\rm{-2}} \sim {\rm{3}}}}$ \cite{NISQ0,NISQ1}, cross-talk noise exists at a level of about 1\%, which, together with the random duration noise modeled above, we refer to as moderate-level noises.

\begin{figure}[h]
  \centering
  \includegraphics[width=0.9\linewidth]{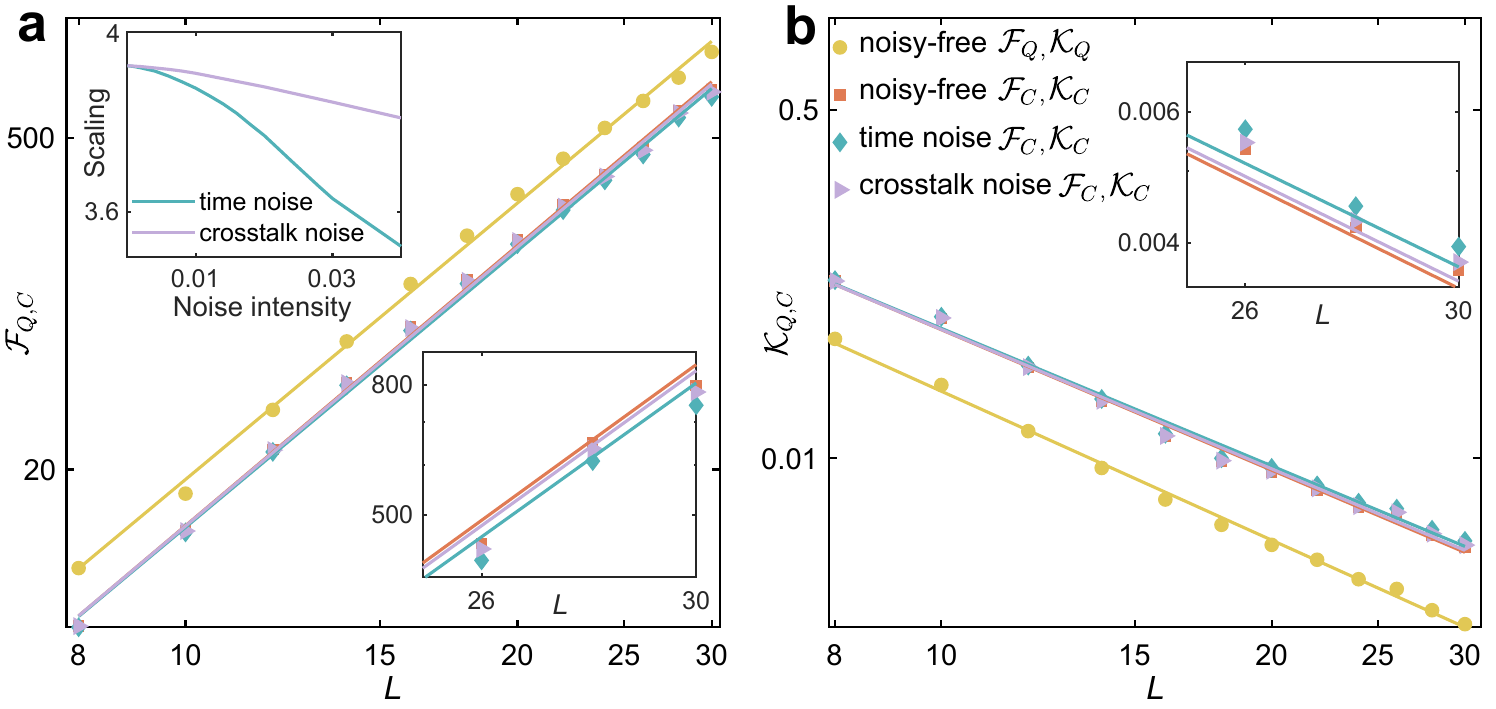}
  \caption{MXY model's sensing performance under optimal single-site measurement of noisy QRLCS for the local (a) and global (b) case.
  The inset in the upper left corner of (a) shows the effect of noise intensity on accuracy limit of QRLCS, where scaling denotes the exponent $a$ in $\mathcal{F_{C}} \propto L^a$.}
  \label{noisyMXY}
\end{figure}

We evaluate the impact of typical noises on the preparation fidelity of the probe. The noises lead to the decrease of preparation fidelity about $2 \sim 10$\% compared to noise-free gates (see SM).
In Fig.~\ref{noisyMXY}, we present the actual accuracy varied with spin length $L$ in the MXY model.
Compared to the noise-free result, these moderate-level noises have small impact on QRLCS protocol.

\textit{Conclusions.}---We propose the QRLCS protocol and demonstrate its effectiveness across various spin models. To address the issue of critical slowing down inherent in conventional methods, we use QRL to prepare probes based on Pauli gates. We then assess the performance of these probes in sensing unknown magnetic fields. Our results conclusively show that the QRLCS protocol achieves the quantum enhanced ($\mathcal{F_{C}} \propto L^{1.62}$) and super-Heisenberg limits ($\mathcal{F_{C}} \propto L^4$) in Ising and XY models, even under Pauli measurements, in both local and global sensing scenarios.
Furthermore, we find that moderate levels of noise have little effect on our protocol. These findings consistently highlight the advantages of our protocol, enabling seamless implementation in real-world applications while maintaining high accuracy.
We deliberate that as quantum computers continue to improve in scale and fault tolerance, the gate fidelity-dependent limitations of QRLCS will be greatly released.
Additionally, our protocol can be easily adapted to other models that utilize different exotic quantum phenomena such as geometric phase transitions \cite{geopt}, thereby broadening its applicability across various sensing scenarios.

\begin{acknowledgments}
This work was supported by the National Natural Science Foundation of China (No. 62401359), the fund of the State Key Laboratory of Advanced Optical Communication Systems and Networks, the Innovation Program for Quantum Science and Technology (No. 2021ZD0300703), Shanghai Municipal Science and Technology Major Project (No. 2019SHZDZX01) and SJTU-Lenovo Collaboration Project (No. 202407SJTU01-LR019).
\end{acknowledgments}

\end{document}